# Maxwell-Hydrodynamic Model for Simulating Nonlinear Terahertz Generation from Plasmonic Metasurfaces

Ming Fang, Zhixiang Huang, *Member*, *IEEE*, Wei E. I. Sha, *Senior Member*, *IEEE*, and Xianliang Wu

*Abstract*—The interaction between electromagnetic field and plasmonic nanostructures leads to both strong linear and nonlinear behaviors. In this paper, a time-domain hydrodynamic model for describing the motion of electrons in plasmonic nanostructures is presented, in which both surface and bulk nonlinearity are considered. A coupled Maxwell-hydrodynamic system capturing full-wave physics and free electron dynamics is numerically solved with the parallel finite-difference time-domain (FDTD) method. The validation of the proposed method is presented by simulating a plasmonic metasurface. The linear response is compared with the Drude dispersion model and the nonlinear terahertz emission from a difference-frequency generation process is theoretically analyzed. The work is fundamentally important to design nonlinear plasmonic nanodevices, especially for efficient and broadband THz emitters.

## I. INTRODUCTION

PLASMONIC nanostructures play an important role in controlling light, for developing photoluminescence enhancement, optical sensing, solar cell, metamaterials and optical antenna [1-5]. These potential applications arise from surface plasmon resonance (SPR) [6], which is induced by the interaction between plasmonic metal nanostructures and external electromagnetic fields. The SPR can be tailored by modifying the geometry and material of nanostructures [7]. Particularly, they are attractive for their inherent nonlinear optical effects. A variety of photonic functionalities based on the nonlinear effects are reported in recent work, such as optical sensing, ultra-short pulse generation, nanoantenna and optical signal processing [8-11]. However, nonlinear interaction in comparison with linear interaction, are intrinsically weak. They are heavily dependent on the field amplitude of excitation. Consequently, high-power excitation is needed to enlarge nonlinear effects. In experiments, an optical laser is adopted for high-power excitation and strong field enhancement can be achieved by carefully designing plasmonic nanostructures. Due to a low conversion efficiency of high-order nonlinearities, the most exciting and useful effects are the second and third-order nonlinearities, such as the second-harmonic generation (SHG), Kerr effect, third-harmonic generation (THG) and four-wave mixing (FWM).

Modeling optical responses of plasmonic nanostructures is a critical step to understand their working principles and perform corresponding designs. For numerical studies, the Drude model is commonly used for describing electron motion in plasmonic materials. However, it is a local-response approximation (LRA) model that does not account for the nonlinear and nonlocal electromagnetic responses. To solve this problem, a classical electrodynamic framework with suitable approximations and improvements can be used. The nonlinear response can be described with material nonlinear polarization, i.e. $\mathbf{P}=\varepsilon_0\chi^{(n)}\mathbf{E}^n$, where $\chi^{(n)}$ is the nonlinear susceptibility [12,13]. Alternatively, a hydrodynamic model in time-domain can be used to model nonlinear effects resulting from motion of conduction electrons in metals. The complex nonlinear dynamics is presented in this model, including the magnetic-field contributions from the Lorenz force, convective acceleration and quantum pressure [14-16]. This hydrodynamic model fully considers the linear and nonlinear dynamics of the electron gas and does not rely on the experimentally measured bulk and surface nonlinear susceptibility [17]. Inspired by the above observations, a time-domain implementation of the hydrodynamic model with the classical computational electromagnetics method is desirable.

The finite-difference time-domain (FDTD) [18] is one of the most commonly-used method to study time-domain behaviors of materials. The FDTD method was originally designed for linear non-dispersive dielectric materials, and has been widely used for more than five decades. With the auxiliary differential equation (ADE) technique [19], the FDTD method is extended to the simulation of dispersive materials. The ADE-FDTD method is flexible for modeling complex permittivity functions and nonlinear effects, such as Kerr and Raman effects in

Manuscript received XX XX, 2017; revised XX XX, 2017; accepted XX XX, 2017. Date of publication XX XX, 2017; date of current version XX XX, 2017. This work was supported the National Natural Science Fund for Excellent Young Scholars (No. 61722101)，NSFC (Nos. 61471001,and 61601166), Universities Natural Science Foundation of Anhui Province (Nos. KJ2017ZD51 and KJ2017ZD02) .

M. Fang, Z. Huang and X. Wu are with the Key Laboratory of Intelligent Computing and Signal Processing, Ministry of Education, Anhui University, Hefei 230039, China (e-mail: zxhuang@ahu.edu.cn)；
W. E. I. Sha is with College of Information Science and Electronic Engineering, Zhejiang University, Hangzhou, 310027, China and the Department of Electrical and Electronic Engineering, The University of Hong Kong. (e-mail: wsha@eee.hku.hk).

nonlinear media [20].

Recently, by using difference-frequency generation, a terahertz (THz) emitter based on a nonlinear metasurface was reported [22]. Efficient, broadband and tunable THz emission can be generated in a thin layer of metamaterial with tens of nanometers thickness by an infrared laser pumping. The THz emission intensity from the metamaterial is on the same order as the nonlinear crystals; and the ultra-thin THz emitter is free from quasi-phase-matching. However, there is no rigorous and universal numerical model to simulate the nonlinear process. In this paper, a self-consistent parallel ADE-FDTD method is employed in conjugation with the hydrodynamic model to simulate the nonlinear THz emission from plasmonic metasurfaces. Compared to the Maxwell-hydrodynamic model in literature [14,23] and our previous work [12,13,24], a new two-step splitting scheme is proposed to solve the multiphysics model. In Ref. [14,16,24], the hydrodynamic equations contain nonlinear terms; the explicit scheme cannot be used directly to solve electron velocity and density, except that the approximations of their values at the previous step are employed. Consequently, both the linear and nonlinear responses are solved with the approximations. To keep the accuracy of the simulation, a fine spatial or temporal increment should then be used. In difference, the proposed ADE-FDTD method uses the two step splitting scheme. The linear effect is explicitly solved by the ADE method and the nonlinear response was explicitly updated with the intermediate value of current obtained from the ADE method. As a result, the linear and nonlinear responses are split and the weak nonlinear signals are not be suppressed by the strong linear pulse signals. The numerical aspects of the proposed model including accuracy, stability and parallel strategy are presented. To demonstrate the performance of the proposed algorithm, the THz signal emitted from a single periodic layer of gold split-ring resonators on glass is investigated. The numerical simulations in a three-dimensional computational domain yield good agreements with the experimental results in [22].

This paper is organized as follows. The numerical scheme and multiphysics model are introduced in Section II. In Section III, the nonlinear response of plasmonic metasurfaces is modeled with the proposed scheme. To benchmark the proposed scheme, the linear transmittance of the metasurface is calculated and the result is compared to the classical Drude model. Then, the THz emission from difference-frequency generation is calculated and compared with theoretical analyses. Finally, this paper is concluded in Section IV.

## II. NUMERICAL SCHEME

### A. The Multiphysics Hydrodynamic Model

The interaction of electromagnetic fields **E** and **H** with nonmagnetic materials is described by the Maxwell's equations

$$\frac{1}{\mu_0}\nabla\times\mathbf{B} = \varepsilon_0 \frac{\partial \mathbf{E}}{\partial t} + \mathbf{J}, \quad (1)$$

$$\frac{\partial \mathbf{B}}{\partial t} = -\nabla\times\mathbf{E}, \quad (2)$$

where $\varepsilon_0$ and $\mu_0$ are the vacuum permittivity and permeability. **J** includes both linear and nonlinear polarization currents, when electromagnetic waves interact with free electron gas in plasmonic materials. The natural dynamics of free electron gas driven by an external electromagnetic wave is described by the hydrodynamic equation

$$\frac{\partial \mathbf{v}_e}{\partial t} + (\mathbf{v}_e \cdot \nabla)\mathbf{v}_e = -\frac{e}{m}(\mathbf{E} + \mathbf{v}_e \times \mathbf{B}) - \gamma \mathbf{v}_e, \quad (3)$$

$$\frac{\partial n_e}{\partial t} + \nabla \cdot (n_e \mathbf{v}_e) = 0, \quad (4)$$

where $m$ and $e$ are the electron mass and electron charge. $n_e$ and $\mathbf{v}_e$ are the electron density and velocity. $\gamma$ is the phenomenological damping frequency constant. To ensure charge neutrality, the initial electron density $n_e$ is set to be equal to positive ion density $n_0$ without excitation. The $(\mathbf{v}_e \cdot \nabla)\mathbf{v}_e$ term is the convective acceleration of electron gas. Equation (4) is the continuity equation. The electromagnetic field and electron dynamics can be coupled via the macroscopic current density and charge density terms

$$\mathbf{J} = -e n_e \mathbf{v}_e, \quad (5)$$

$$\rho = e(n_e - n_0) \quad (6)$$

Substituting (5) and (6) into (3) and (4), we have

$$\frac{\partial \mathbf{J}}{\partial t} = \varepsilon_0 \omega_p^2 \mathbf{E} - \gamma \mathbf{J} + \frac{e}{m}(\rho \mathbf{E} - \mathbf{J}\times\mathbf{B}) + \nabla\cdot\left(\frac{1}{\rho + \varepsilon_0 m \omega_p^2 / e}\mathbf{JJ}\right) \quad (7)$$

$$\frac{\partial \rho}{\partial t} - \nabla\cdot\mathbf{J} = 0, \quad (8)$$

where $\omega_p = \sqrt{e^2 n_0 / \varepsilon_0 m}$ is the plasma frequency. Finally, Eqs. (7) and (8) can be reduced to one equation by substituting $\rho = \varepsilon_0 \nabla\cdot\mathbf{E}$ into them

$$\frac{\partial \mathbf{J}}{\partial t} = \varepsilon_0 \omega_p^2 \mathbf{E} - \gamma \mathbf{J} + \frac{e}{m}\left[\varepsilon_0(\nabla\cdot\mathbf{E})\mathbf{E} - \mathbf{J}\times\mathbf{B}\right] \quad (9)$$
$$+ \nabla\cdot\left(\frac{1}{\varepsilon_0(\nabla\cdot\mathbf{E}) + \varepsilon_0 m \omega_p^2 / e}\mathbf{JJ}\right)$$

Equations (1), (2) and (9) form the hydrodynamic model for linear and nonlinear interaction between the electromagnetic wave and free electron gas in plasmonic nanostructures.

### B. Computational Grid

In order to solve (1),(2) and (9), a computational grid based on the standard staggered Yee cells is proposed, as seen in Fig.

1. The spatial-temporal dependent **E**, **H** and **J** are nodally uncollocated in space and staggered in time. The electric field **E** is defined at time step $l+1/2$ and is located at the cell face center. Both current density **J** and magnetic field **H** are defined at the time level $l$ and are located at the cell face centers and the edges, respectively. This grid arrangement captures the properties of fields and charges as well as the cross-coupling effects between free electrons and electromagnetic fields.

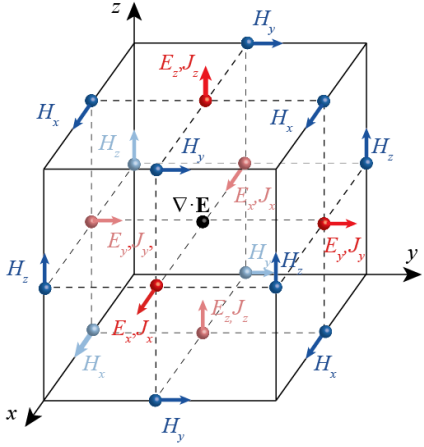

Fig. 1. The Yee grids for the Maxwell-hydrodynamic system

### C. Two-step Splitting Scheme

We employ an explicit central-difference method to discretize Eqs. (1) and (2). Considering the sampled time level of **J** and the nonlinear properties of Eq. (9), a time-splitting scheme is applied for updating **J**. We split Eq. (9) into two subproblems

$$\frac{\partial \mathbf{J}}{\partial t} = \varepsilon_0 \omega_p^2 \mathbf{E} - \gamma \mathbf{J} \quad (10)$$

$$\frac{\partial \mathbf{J}}{\partial t} = \frac{e}{m}\left[\varepsilon_0(\nabla \cdot \mathbf{E})\mathbf{E} - \mathbf{J}\times\mathbf{B}\right] + \nabla \cdot \left(\frac{1}{\varepsilon_0(\nabla \cdot \mathbf{E}) + \varepsilon_0 m\omega_p^2/e}\mathbf{JJ}\right) \quad (11)$$

The first equation (10) is the linear response of electrons as described by the conventional Drude model. Updating **J** from the time step $l$ by a central-difference scheme, we obtain an intermediate updated current density $\mathbf{J}^{(1)}$ at time step $l+1$,

$$\mathbf{J}^{(1),l+1} = \frac{1-0.5\Delta t\gamma}{1+0.5\Delta t\gamma}\mathbf{J}^l + \frac{\varepsilon_0 \omega_p^2 \Delta t}{1+0.5\Delta t\gamma}\mathbf{E}^{l+1/2}. \quad (12)$$

Equations (11) present the nonlinear response. Since current density **J** is defined at the same time level as **H** and is at the same location as **E** in space. The $\varepsilon_0 \nabla \cdot \mathbf{E}$ term in Eq. (11) representing the charge density is first calculated at the center of grid. The components are interpolated to approximate the values at the same spatial and temporal point as **J**. For example, for updating the $x$-component of **J**, the interpolated $\nabla \cdot \mathbf{E}$ and **B** are given by

$$\overline{\nabla \cdot \mathbf{E}}_{i,j+1/2,k+1/2}^{l+1/2} = \frac{1}{2}(\nabla \cdot \mathbf{E}_{i+1/2,j+1/2,k+1/2}^{l+1/2} + \nabla \cdot \mathbf{E}_{i-1/2,j+1/2,k+1/2}^{l+1/2}), \quad (13)$$

$$\overline{B}_{y\,i,j+1/2,k+1/2}^{l+1/2} = \frac{1}{4}(B_{y\,i,j+1/2,k}^{l+1} + B_{y\,i,j+1/2,k+1}^{l+1} + B_{y\,i,j+1/2,k}^{l} + B_{y\,i,j+1/2,k+1}^{l}) \quad (14)$$

$$\overline{B}_{z\,i,j+1/2,k+1/2}^{l+1/2} = \frac{1}{4}(B_{z\,i,j,k+1/2}^{l+1} + B_{z\,i,j+1,k+1/2}^{l+1} + B_{z\,i,j,k+1/2}^{l} + B_{z\,i,j+1,k+1/2}^{l}) \quad (15)$$

Meanwhile, **J** is interpolated at the time level $l+1/2$ by averaging $\mathbf{J}^l$ and the intermediate current density $\mathbf{J}^{(1),l+1}$

$$\overline{J}_{x\,i,j+1/2,k+1/2}^{l+1/2} = \frac{1}{2}(J_{x\,i,j+1/2,k+1/2}^l + J_{x\,i,j+1/2,k+1/2}^{(1),l+1}) \quad (16)$$

After explicitly solving Eq. (11), the nonlinear intermediate current density $\mathbf{J}^{(2)}$ can be obtained. After getting the second intermediate value of $\mathbf{J}^{(2)}$, a final current density at the time step $l+1/2$ is given by $\mathbf{J}=\mathbf{J}^{(1)}+\mathbf{J}^{(2)}$ for updating the next step electric field. Instead of solving a large matrix with an implicit difference method, the equation of hydrodynamic model is split into a linear explicit equation and a nonlinear explicit equation. The proposed two-step splitting scheme is adopted for efficiency and simplicity of the implementation.

### D. Numerical Stability

The stability of the proposed scheme can be analyzed by finding the roots of the corresponding growth matrix. Unfortunately, even with modern computer techniques, expressions for the roots are too complex to obtain. Here, the nonlinear terms in the hyperbolic equation Eq. (15) mainly contribute to the high-order harmonic generation. Compared with the linear term in Eq. (13), the nonlinear responses are small. After modeling different complex metallic nanostructures for a long-term simulation, we found that the hydrodynamic model was stable within the commonly used CFL constraint of the FDTD method. In other words, the nonlinear terms, which are much weaker than the linear one, do not significantly affect the stability condition of the FDTD method with the linear Drude model.

### E. Parallel strategy

The plasmonic nanodevices, such as metamaterials, are subwavelength nanostructures exhibiting a strong near-field enhancement. Simulations of them need a great amount of grids to achieve a high spatial resolution. Meanwhile, the bandwidth of the fundamental pulse should be narrow enough in the frequency domain so that the high-order harmonic signals are not overwhelmed by the fundamental one. As a result, long CPU time with large iteration steps are required to guarantee the sufficient decay of the fundamental pulse. To fix this problem, a parallel computing technique can be employed. Thanks to the localization nature of the FDTD method, we can divide the whole computational domain into many sub-domains.

A high efficient parallel code can then be achieved by implementing the FDTD code on computer cluster.

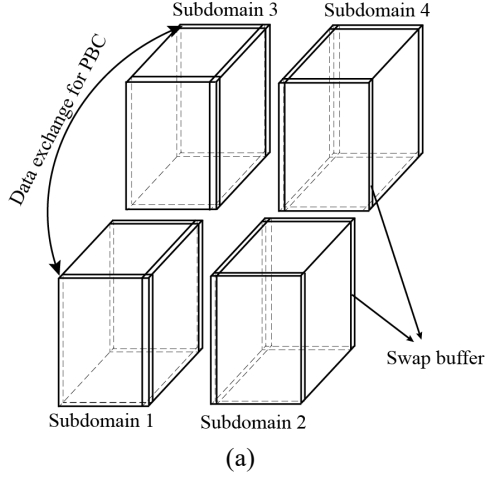

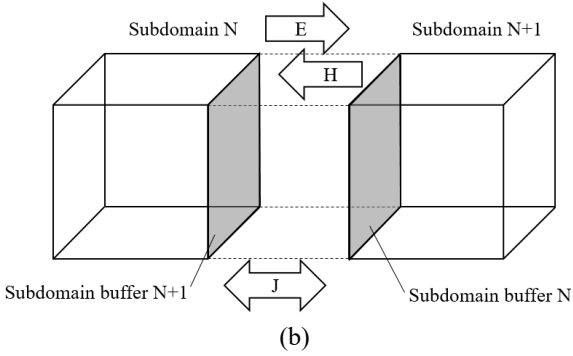

Fig. 2. Diagram of the parallel computing scheme. (a) The parallel FDTD processing on a computer cluster. (b) field exchange between the adjacent subdomains.

Fig. 2 shows our parallel FDTD algorithm. The simulation domain is divided into many subdomains and each subdomain is assigned to a processor. Each subdomain executes the same code in parallel and swaps the data at the interface at every time step. The data exchange between different processors or domains is carried out by using the message passing interface (MPI) library. At the subdomain boundaries, one additional layer of cells is allocated as a swap buffer. The flowchart of the subdomain is diagrammed in Fig. 3. Comparing with the Drude model, the field iterations of the hydrodynamic model needs the fields in the neighboring Yee grids, thus both the components **E**, **H** and **J** are stored in the swap buffer and exchanged at the interface. In addition, the data exchange is also needed for the periodic boundary condition.

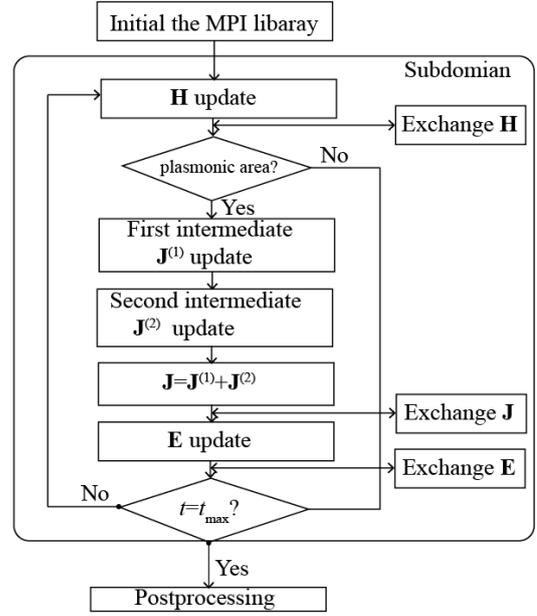

Fig. 3. The flowchart of a subdomain in the parallel FDTD scheme.

## III. NUMERICAL VALIDATION AND APPLICATION

The nonlinearity of plasmonic nanodevices is simulated in this section. This problem is motivated by a recent experiment research [22]. A broadband and efficient THz emission from a plasmonic metasurface, based on resonant photoexcitation of the magnetic dipole resonance. The THz emission from the metamaterials is continuous in the THz regime and its intensity is on the same order as optimal crystals that are thousand times thicker.

### A. Configurations

In a recent work, researchers have shown that a THz source ranging from 0.1 THz to 4 THz can be generated by a plasmonic metasurface [22]. However, due to the upper cut-off frequency of the inorganic crystal detector is much smaller than the emitted THz signal, the measured bandwidth of THz signal was much narrower than the expected. Therefore it is important to investigate the mechanism of the THz emission from metasurface by a numerical method. The efficient and compact metasurface THz emitter is shown in Fig. 4 (a). The gold SRRs are arranged periodically in both *x* and *y* directions and supported by a layer of glass substrate coated with ITO. The thickness of the gold film, ITO and $SiO_2$ substrates are 40 nm, 6 nm and 200 nm, respectively. The glass substrate and ITO layer are chosen as nondispersive dielectric materials and the relative permittivity are 2.25 and 3.8, respectively. The parameters for gold are taken to be $n_0 = 5.92 \times 10^{28}$ m$^{-3}$ and $\gamma = 10.68 \times 10^{13}$ rad/s. The dimensions of the SRRs are chosen as shown in Fig. 4 (b).

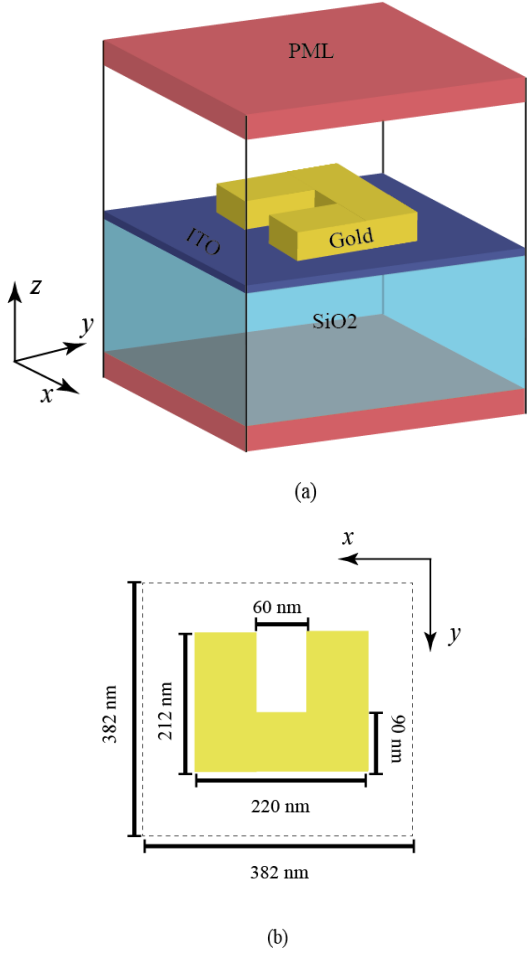

(a)

(b)

Fig. 4. (a) Layout of a metasurface consist of a periodic array of SRRs. (b) The feature size of a unit cell.

*B. Validation with linear response*

We firstly consider the validation for the linear response of our method. The set of Eqs. (1), (2), (10) and (11) are numerically solved by employing the ADE-FDTD method. A plane wave source is introduced by using the total-field and scattered-field (TF/SF) technique and it propagates in the z-direction. PMLs are set to be at the *z* direction and periodic boundary conditions are employed at both the *x* and *y* directions. Uniform spatial steps $\Delta x = \Delta y = \Delta z = 2$ nm and a time step $\Delta t = 3\times10^{-18}$ s are applied. A computational domain of $190\times190\times1000$ grids is used. The calculated transmittance spectrum of the plasmonic metasurface around the fundamental magnetic resonance frequency is calculated numerically by using the hydrodynamic model and compared with the Drude model as shown in Fig. 5 (a). The solid line indicates the spectra obtained by the hydrodynamic model and the dashed line is the spectra calculated by the Drude model. Fig. 5 (b) shows the relative errors. One can see that the results obtained using our proposed method are in good agreements with those obtained with the classical Drude method.

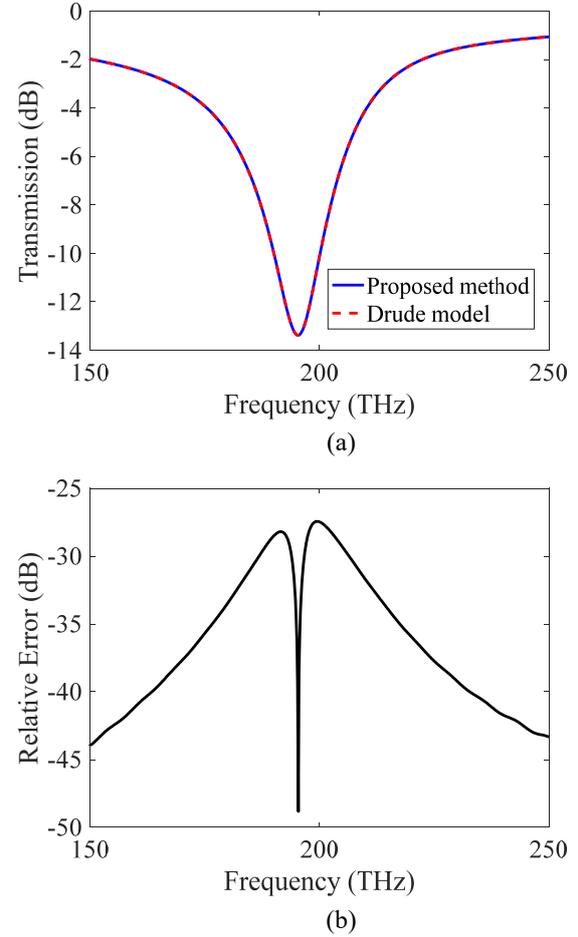

Fig. 5. (a) The calculated transmittance of the metasurface shown in Fig. 4. The results are shown for *x*-polarized excitation. (b) Relative errors between our method and the Drude model.

*C. Nonlinear simulation*

The THz emission from the metasurface due to optical rectification (OR) or difference-frequency generation is induced by the symmetry breaking in the second order nonlinear current distribution. From Eq. (9), we can see that the convective acceleration term $(\mathbf{J}\cdot\nabla)\mathbf{J}$ and $(\nabla\cdot\mathbf{E})\mathbf{E}$ are the key contributions to the second-order nonlinearities. These terms behave like linear current times linear charge density $\mathbf{J}\rho$ consequently, the THz nonlinear current is parallel or antiparallel to the linear current. When the polarization of incident wave is parallel to the gap of SRRs (*x*-polarization), the nonlinear currents at both arms are parallel and radiate fields that can be observed in the far field.

To validate the nonlinear response of our method, we model the THz emission from the metasurface that is excited by an *x*-polarized laser. The pump laser is a Gaussian pulse

$$E(t) = \cos(\omega_0 t + \phi)g(t), \quad (17)$$

where

$$g_\sigma(t) = e^{-\frac{1}{2}\sigma^2 t^2} \Leftrightarrow g_\sigma(\omega) = \frac{\sqrt{2\pi}}{\sigma}e^{-\frac{\omega^2}{2\sigma^2}} \quad (18)$$

is a Gaussian envelope. $\omega_0$ is the center frequency of the pump pulse and $\phi$ is a phase delay. The second-order nonlinear polarization from the plasmonic material is proportional to the power of the incident pump pulse

$$\begin{aligned} P^{(2)}(t) &\sim E^{(2)}(t) \\ &= \cos(2\omega_0 t + 2\phi)g_\sigma^2(t) + g_\sigma^2(t) \end{aligned}, \quad (19)$$

$$\begin{aligned} P^{(2)}(\omega) &\sim e^{i2\varphi}g_{\sqrt{2}\sigma}(\omega - 2\omega_0) \\ &+ e^{-i2\varphi}g_{\sqrt{2}\sigma}(\omega + 2\omega_0) + g_{\sqrt{2}\sigma}(\omega) \end{aligned}. \quad (20)$$

Here, the first two terms are corresponding to the second nonlinearity with the center frequency $2\omega_0$; the last term is the THz emission; thus, the THz emission only depends on the temporal Gaussian envelope. The emitted THz field has an analytical solution given by

$$\begin{aligned} E^{THz}(t) &\sim \partial^2 P / \partial t^2 \sim \chi^{(2)}\partial_t^2 g_{\sqrt{2}\sigma}(t) \\ &= \chi^{(2)}\sigma^2(1 - 2\sigma^2 t^2)e^{-\sigma^2 t^2} \end{aligned} \quad (21)$$

The linear and nonlinear signals are calculated numerically with the proposed method. The driving frequency, temporal width, change rate and peak amplitude of the pump laser are chosen as $\omega_0 = 1.228\times10^{15}$ rad/s, $\sigma = 2\ln 2/\tau$ and $E_0 = 2\times10^7$ V/m, respectively. Here $\tau = 150$ fs is the duration of excitation pulse corresponding to a spectral width of 3 THz. The total number of simulation time steps is $4.0\times10^5$ corresponding to a physical time of 1.2 ps. Fig. 6 shows the temporal profile of linear (a) and nonlinear transmitted field (b). The linear and nonlinear spectra are plotted in Fig. 6 (c) after the Fourier transform. We can see that the second-harmonic and difference-frequency generation are observed clearly.

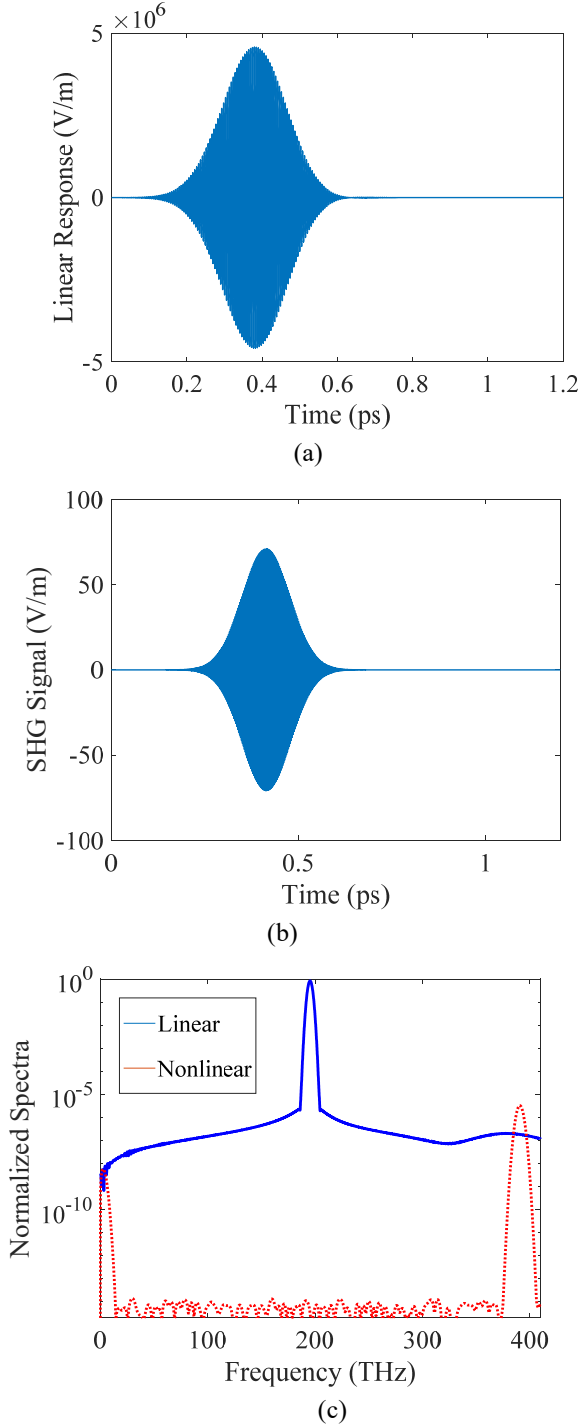

Fig. 6. Electric field as a function of time for the linear (a) and nonlinear (b) transmission. (c) Semi-log plot of the Fourier transform of the linear and nonlinear transmitted fields versus frequency.

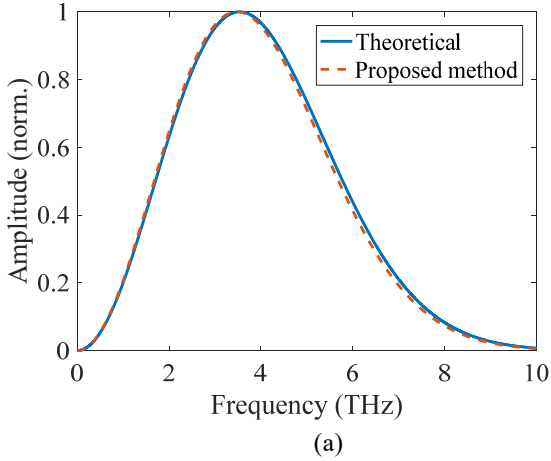

(a)

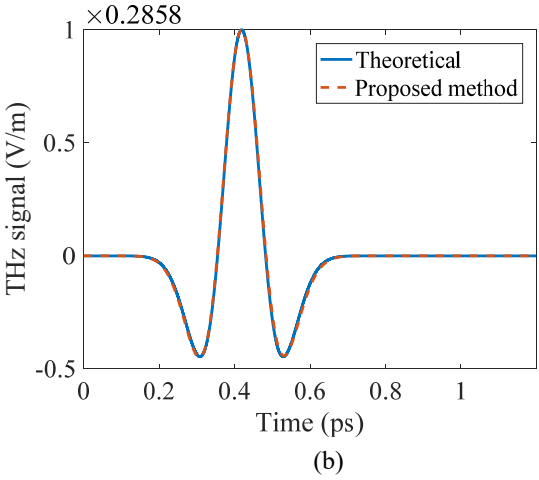

(b)

Fig. 7 The normalized THz spectrum (a) and corresponding time-domain THz pulse (b).

The THz signals in the frequency domain are extracted. First a Fourier transform is performed on the time-domain nonlinear signals, in which a specific frequency range of 0~10 THz is selected while the intensity of all other frequencies outside the selected window is set to zero. Then we transform this windowed frequency-domain signals back into its time-domain counterparts. Since the resonance of SRRs is much wider (35.2 THz shown in Fig. 5(a)) compared to the pump pulse width, and thus the emitted THz bandwidth from SRRs is limited by the bandwidth of the incident pump pulse. As a comparison, the predicted result of analytical solution is plotted in Fig. 7, from which excellent agreement can be observed. Note that the THz emission bandwidth of our results is broader than the observed THz spectrum in the experiment [22]. This is because the upper cut-off frequency of the THz signal detector in the experiment is smaller than the theoretical prediction. The metasurface THz emitter can generate broadband THz signals by shortening the pump pulse duration because the single-layer emitter of 40 nm thickness is continuous and free from the quasi-phase-matching limitation at the THz region.

By using the parallel strategy in previous section, the performance of our parallel FDTD code can be improved. For the same configuration of 190×190×400 grid, the computation speedup versus the number of processors is plotted in Fig. 8. The computations are performed on a workstation with 4 Intel(R) Xeon(R) E5-2680 v2 @ 2.80GHz CPU (a total of 40 CPU cores). The total computational memory was about 5.4 GB and total computational time was about 20 hours with 32 processors. The practical speedup is lower than theoretically expected. This is due to the increased data transfers (**E**, **H**, **J**) between processors. In addition, the implementation in the plasmonic material area is complicated, the simulation time is much large for the subdomain with plasmonic structures. Therefore, simulations of the subdomains decrease the parallel computational efficiency.

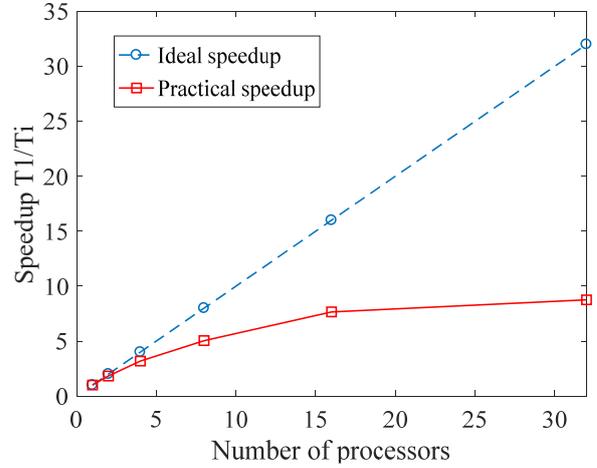

Fig. 7 Computational speedup versus number of processors.

### IV. CONCLUSIONS

In conclusion, a self-consistent, accurate, and explicit FDTD based Maxwell-hydrodynamic model has been developed for investigating and modeling both linear and nonlinear electromagnetic responses in plasmonic metasurfaces. In particular, a time-splitting scheme is adopted for solving the hydrodynamic equation. With the proposed method, the linear transmittance of a plasmonic metasurface is simulated and compared with the Drude model, and the emitted nonlinear THz signal is accurately solved. Consequently, the proposed multiphysics FDTD scheme present a way for understanding and designing nonlinear plasmonic nanodevices.

ACKNOWLEDGMENT

M. Fang would like to thank C. M. Soukoulis and Th. Koschny, for their help and guide at Iowa State University, USA.